\documentclass{iopart}
\usepackage{graphicx}
\usepackage{cite}
\usepackage{iopams}
\newcommand{\dr}{{\rm d}}

\newcommand{\om}{\omega}
\newcommand{\Om}{\Omega}
\newcommand{\ep}{\epsilon}
\newcommand{\de}{\delta}
\newcommand{\bea}{\begin{eqnarray}}
\newcommand{\beq}{\begin{equation}}
\newcommand{\eea}{\end{eqnarray}}
\newcommand{\eeq}{\end{equation}}
\begin{document}

\title[Fano-Feshbach resonances in two-channel scattering around exceptional points]
{Fano-Feshbach resonances in two-channel scattering around exceptional points}

\author{W D Heiss$^{1, 2}$, G Wunner$^{1, 3}$}
\address{$^1$Department of Physics, University of Stellenbosch,
  7602 Matieland, South Africa}
\address{$^2$National Institute for Theoretical Physics (NITheP), Western Cape,\
  South Africa}
\address{$^3$Institut f\"ur Theoretische Physik, Universit\"at Stuttgart,
  Pfaffenwaldring 57, 70\,569 Stuttgart, Germany}

\ead{dieter@physics.sun.ac.za}

\begin{abstract}
It is well known that in open quantum systems resonances can coalesce at an exceptional point, 
where both the energies  {\em and} the wave functions coincide. In contrast to the usual 
behaviour of the scattering amplitude at one resonance, the coalescence of two resonances invokes
a pole of second order in the Green's function, in addition to the usual first order pole. 
We show that the interference due to the two pole terms of different order gives rise 
to patterns in the scattering cross section which closely resemble Fano-Feshbach resonances. We demonstrate this 
by extending previous work on the analogy of Fano-Feshbach resonances to classical resonances in a system of
two driven coupled damped harmonic oscillators. 
\end{abstract}

% Uncomment for PACS numbers title message
% \vspace{2pc}
% \noindent{\it Keywords}: Article preparation, IOP journals
\submitto{\JPA}
\maketitle

\section{\label{sec:intro} Introduction}
When a quantum state in the continuum interacts with another discrete state, 
the two states give rise to a resonance in the continuum channel showing an asymmetric line profile in the cross section. 
The theoretical description of these resonances was developed independently by Feshbach \cite{Feshbach58} in nuclear physics using projection operator techniques, and by Fano \cite{Fano61} in atomic physics using the language of superposition of wave functions. These Fano-Feshbach resonances are ubiquitous in physics, and have
more recently also been found in mesoscopic condensed matter systems \cite{Rau04} and nanoscale structures \cite{Miroshnichenko10}). They have also been used in Bose-Einstein  condensates to tune the scattering interaction \cite{Chin10}. Joe et al. \cite{Joe06} have pointed out that one can even draw an analogy between quantum interference of Fano-Feshbach
resonances and classical resonances in a system of two driven coupled damped oscillators;
one of the oscillators exhibits the standard enhancement of the amplitude near its eigenfrequency, while near the
eigenfrequency of the second oscillator the amplitude acquires an asymmetric profile, and at a certain
frequency of the driving force it can even become zero. Effectively there are two driving
forces acting on the first oscillator which are out of phase and may cancel each other. This example demonstrates
one of the basic features of the Feshbach-Fano resonance, namely resonant destructive interference.

The same classical system has also been analyzed \cite{Heiss04} to exemplify typical behaviour of quantum systems
with non-Hermitian operators. These operators occur for example in open quantum systems, which are in contact with an environment, and where resonances with complex energies appear. Other examples of systems which can
be described by non-Hermitian operators can be found, e.~g., in Ref.~\cite{Moiseyev2011a}. It is well known that
resonances show characteristic effects not observable in Hermitian quantum systems.  Among these effects
are exceptional points (EPs) \cite{Kato66} being isolated points in an (at least) two-dimensional 
parameter space at which two or even more eigenstates coalesce. EPs exhibit distinct features
and usually influence in their vicinity the behaviour in parameter space in a specific way \cite{Heiss10a}. They show  particular
properties such as the permutation of eigenstates for a closed adiabatic loop in parameter space \cite{Maily05},
a special geometric phase {for the one only eigenstate at the EP \cite{Heiss99}, or a linear term in the time evolution of the wave function besides
the usual exponential behaviour \cite{Heiss10b}. EPs feature in quantum mechanical as well 
as in classical systems, examples can be found in Ref. \cite{Heiss12}. 

A further important property of EPs is the appearance of a pole of second
order in the Green's function \cite{Hernandez00,Fuchs14} when an EP is approached in parameter space; 
this second order pole occurs in addition to the pole of first order usually associated with resonances. It is the purpose
of this paper to demonstrate how the approach of these poles of first order and their merger with an additional pole of second order impacts on the scattering cross section. We demonstrate that it is
even already in the vicinity of the merging  of the two poles where the specific Fano-Feshbach-like features in the cross section occur.
As shown in Refs. \cite{Heiss01} the generic behaviour of systems in the vicinity of an EP can be stripped down
to a two-dimensional matrix eigenvalue problem, irrespective of the concrete classical or quantum system under
consideration. In the present paper we use the above mentioned 
driven classical system of two coupled oscillators with damping to mimic the general behaviour of 
a two-channel $T$ matrix in open quantum systems when an EP is approached; 
note that the open system implies non-unitarity of $T$.
 In Sect.~2 we briefly review the classical problem. In Sect.~3 we present the results and 
relate them to the ongoing literature. In  Sect.~4 we draw conclusions and give an outlook.
 
%\cite{Hernandez00}, \cite{Heiss04}, \cite{Stehmann04}, \cite{Heiss10a}, \cite{Heiss10b}, \cite{Mostafazadeh09}, %\cite{Mostafazadeh13a}, \cite{Mostafazadeh13b}, 
%\cite{Miroshnichenko10}, \cite{Joe06}, \cite{Kobayashi02}, 

\section{The model}
In line with the previous discussion,  and for the reader's convenience, we briefly review the model of two coupled classical oscillators \cite{Heiss04}. 
Denoting by $p_1,p_2,q_1,q_2$ the momenta
and spatial coordinates of two point particles of equal mass the equations of
motion read for the driven system
\beq
{\dr \over \dr t}\pmatrix{p_1 \cr p_2 \cr q_1 \cr q_2}={\cal M}
\pmatrix{p_1 \cr p_2 \cr q_1 \cr q_2} 
 + \pmatrix{c_1 \cr c_2 \cr 0 \cr 0}\exp (\rmi\omega t)
\label{eom}
\eeq
with 
\beq
{\cal M}=\pmatrix{-2g -2k_1 & 2g & -f-\omega_1^2 & f \cr
2g & -2g -2k_2 & f & -f-\omega_2^2 \cr
1 & 0 & 0 & 0 \cr
0 & 1 & 0 & 0 } 
\label{mat}
\eeq
where $\tilde \omega _j - \rmi k_j,\; j=1,2$, with $\tilde \omega_j=\sqrt{\omega_j^2-k_j^2}$, 
are the unperturbed damped frequencies while
$f$ and $g$ are the coupling spring constant and damping of the coupling,
respectively. The driving force is assumed to be oscillatory with one single
frequency and acting on each particle with amplitude $c_j$. 
Here we are interested only in the stationary solution being the
solution of the inhomogeneous equation which reads
\beq
\pmatrix{p_1 \cr p_2 \cr q_1 \cr q_2}=(\rmi\omega -{\cal M})^{-1}
\pmatrix{c_1 \cr c_2 \cr 0 \cr 0}\exp (\rmi\omega t).
\label{sol}
\eeq
Resonances occur for the real values $\omega $ of the complex solutions of the
secular equation
\beq
\det |\rmi\omega -{\cal M}|=0
\label{det}
\eeq
and EPs occur for the complex values $\omega $ where
\beq
{\dr \over \dr \omega}\det |\rmi\omega -{\cal M}|=0
\label{der}
\eeq
is fulfilled simultaneously together with Eq.(\ref{det}). We choose the
parameter $f$ as the second variable needed to enforce the simultaneous solution
of Eqs.(\ref{det}) and (\ref{der}) and keep the other parameters of ${\cal M}$ fixed.
Thus we encounter the problem of finding
the EPs of the matrix problem
\beq
{\cal M}_0+f {\cal M}_1  \label{fep}
\eeq
with
\beq
{\cal M}_0=\pmatrix{-2g -2k_1 & 2g & -\omega_1^2 & 0 \cr
2g & -2g -2k_2 & 0 & -\omega_2^2 \cr
1 & 0 & 0 & 0 \cr
0 & 1 & 0 & 0 } 
\eeq
and
\beq 
{\cal M}_1=\pmatrix{0&0&-1&1\cr 0&0&1&-1
\cr 0&0&0&0 \cr 0&0&0&0 }.
\eeq
Note that ${\cal M}_0$ and ${\cal M}_1$ are not symmetric. In the numerical search
for the EP we enforce $f$ to be real. This is achieved by adjusting accordingly the value of
the real parameter $g$. In other words, instead of the two parameters of a complex $f$ we
choose the two real values for $f$ and $g$. This choice is of course motivated on physical grounds.
The corresponding frequency values $\omega _{\rm EP}$
will always be complex irrespective of the choice (complex or real) of the unperturbed frequencies
$\tilde \omega _j - \rmi k_j$. For physical reasons we choose an EP with $\Im \omega _{\rm EP} < 0$.

In the neighbourhood of the EP we reduce the actual problem to an effective two-channel
scattering problem. It is well known that a higher dimensional eigenvalue problem can be reduced
to a two-dimensional matrix problem in close vicinity of an EP \cite{Heiss01}. The advantage lies in the
analytic availability of the Green's function and thus of the scattering amplitudes. It enables
the understanding of the effects of the EP upon the pattern of the scattering at and in close vicinity
of an EP. In fact, as mentioned above the coalescence of two resonance poles -- i.e.~at the EP -- leads to a pole of second
order \cite{Hernandez00,Fuchs14,Heiss10a} besides the usual first order pole. This is made explicit in 
the reduced matrix which can be written in the form
\bea
H(f)&=& H_0 +f V \nonumber\\
&=&
\pmatrix{\Om_1 & 0 \cr 0 & \Om_2 }
+f \pmatrix{\ep_1 & \de_1 \cr \de_2 & \ep_2}
\label{ham}
\eea
where all entries are obtained numerically from the original physical parameters in Eq.(\ref{mat}). One finds the EPs at
\beq
f_{\rm EP}=\frac {-\rmi (\Om _1-\Om _2)}{ \rmi (\ep _1-\ep _2)\pm 2 \sqrt{\de_1 \de_2} }
\label{EP}
\eeq
with the Green's function  
\bea
G(E)\quad =\quad (E-H(f_{\rm EP}))^{-1} &=& \nonumber \\
\frac {1}{E-\om(f_{\rm EP})}\pmatrix{1 & 0 \cr 0 & 1} +
\frac {\rmi \sqrt{\de_1\de_2}f_{\rm EP}}{(E-\om(f_{\rm EP}))^2}&&\pmatrix{1 & \rmi \sqrt{\frac{\de_1}{\de_2}} \cr
\rmi \sqrt{\frac{\de_2}{\de_1}} & -1}.
\label{Green}
\eea
The explicit form for $\om(f_{\rm EP})$ can be found in \cite{Heiss12}.
We emphasise that the values of the EPs of the matrices in
Eq.(\ref{ham}) and Eq.(\ref{fep}) coincide and so do the energy trajectories
in the vicinity of the EPs.

When calculating the scattering amplitudes $T(E)=V+VG(E)V$ and a cross section from $|T_{1,1}|^2$ or
$|T_{2,2}|^2$ the
interference term between the second and first order pole terms turns out to be of crucial importance.
Moreover, even when moving slightly away from the EP by choosing $f$ somewhat different from $f_{\rm EP}$
the interference of the two resonance poles appearing now yields two peaks in $|T_{1,1}|^2$
or $|T_{2,2}|^2$ that cannot be identified with either individual pole term. Quantitative results of this striking
phenomenon will be discussed in the following section.

\section{Results and discussion}
For a first discussion we use the parameter set $\om_1=2.8,\, \om_2=3,\, k_1=k_2=0 $.
A physically acceptable EP is found at $\om_{\rm EP}=2.9-\rmi \, 0.1$ for $f_{\rm EP}=0.02$
and $g_{\rm EP}=0.1$ (the actual numerical values are obtained at much higher precision).
Using these values the plot of $|T_{2,2}|^2$ {\it versus} the energy is shown on top of Fig.1
(the corresponding plot for $T_{1,1}$ looks similar). The distinct two peaks are {\it not} related to an individual
pole of first or second order, the whole pattern is rather brought about by the sum of
the moduli squared of the two pole terms plus the interference term; in particular, the zero of the scattering
function between the two peaks is invoked by the interference term. To illustrate
the dramatic effect we display on the bottom of Fig.1 the interference term; note
the difference in scale by an order of magnitude. Of course the contributions from the individual pole terms --
at the same position -- are of similar order of magnitude.

The obvious follow-up question is: how does the pattern change under a small
variation of, say, $f$? When $f$ is moved away from $f_{\rm EP}$, two individual
resonance pole terms emerge from the EP, one with a larger width (imaginary part) and the
other with a smaller width. In fact, for $f=f_{\rm EP}+0.2$, the pole moving to the left
has diminished its width while the width of the pole moving to the right has grown larger.
For $f=f_{\rm EP}-0.2$ it is just the other way around. The illustration in Fig.2 where the two
cases are displayed seems to confirm this by comparing the widths of the peaks of the left
and the right figure. However, it would be fallacious to associate either of the two
peaks with one of the two resonance poles as the interference term still has its notable
effect. The two poles without interference could not even be resolved as they would be two
overlapping resonances. Rather, not only the zero between the peaks but also their width
and position is effected by the interference term. This is illustrated in Fig.3 where
all terms are illustrated for the two cases.

\begin{figure}[tb]
\begin{center}
\includegraphics[width=0.4\textwidth]{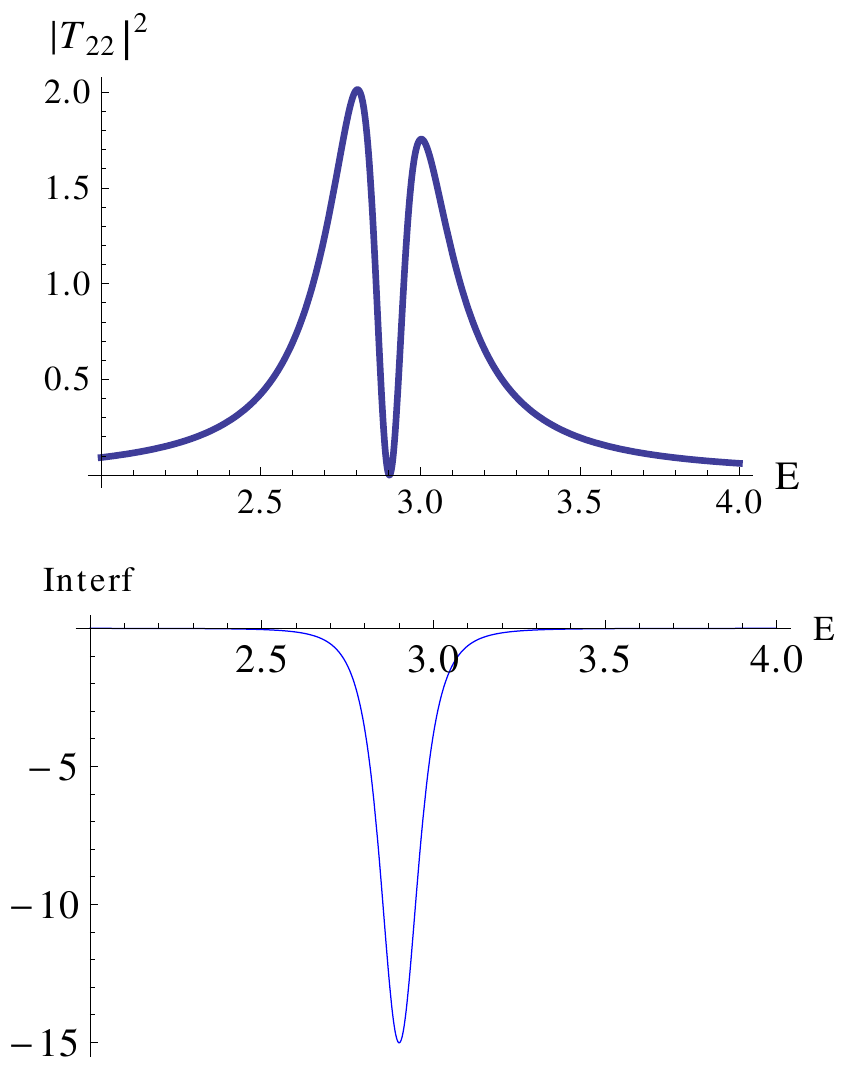}
\end{center}
%\begin{minipage}[b]{12pc}
\caption{Top: Cross section $|T_{22}|^2$ in arbitrary units at $f=f_{\rm EP}$ {\it vs.} energy;
bottom: interference term, i.e.~$ 2\Re[T^{(2)}_{22}\,(T^{(1)}_{22})^*]$ where
$T^{(2)}_{22}$ and $T^{(1)}_{22}$ are the contributions from the second and first order
pole terms, respectively; same units as on top.}                  
%\end{minipage}
\end{figure}

Moving even further away with $f$ from $f_{\rm EP}$, the scattering gradually assumes
a form where each peak can be associated with one resonance pole as the interference term
has become small in comparison with the pole terms. And yet, for our parameters the interference term still
forces the vanishing of scattering between the two peaks. The illustrations in Fig.4
for $f=f_{\rm EP}\pm 1$ are to clarify this point. We mention that the zero, meaning no scattering
at this energy point, has also been observed in a similar setting for complex scattering
potentials modelling wave guides with $\cal PT$-symmetric
gain and loss \cite{Mostafazadeh09,Mostafazadeh13a}.

\begin{figure}[tb]
\begin{center}
\includegraphics[width=0.5\textwidth]{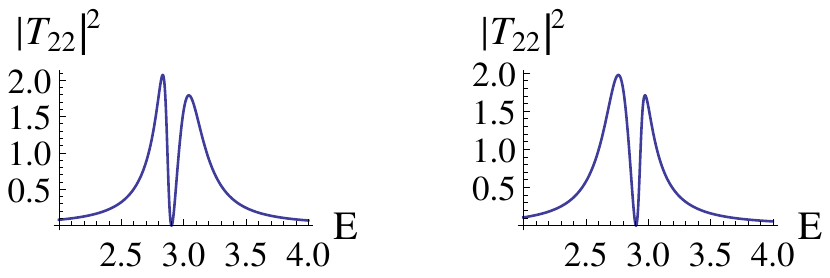}
\end{center}
%\begin{minipage}[b]{12pc}
\caption{Cross sections $|T_{22}|^2$ in arbitrary units {\it vs.} energy
for $f=f_{\rm EP}+0.2$ (left) and $f=f_{\rm EP}-0.2$ (right).  }                  
%\end{minipage}
\end{figure}

When instead of changing $f$ other parameters are changed, the pattern is qualitatively
similar. One notable difference is caused by giving the unperturbed frequencies a finite
width denoted above by $k_1$ and $k_2$. If either one or both  are different from zero the
zero between the peaks of the scattering function disappears, there remains a dip without
touching the zero line. A similar behaviour has been observed by Joe et al. \cite{Joe06} for the amplitudes
of their oscillator system. In actual experiments showing Fano-Feshbach resonances this is 
the rule. 
There are in fact experiments where the asymmetry 
and the depth of the dip can be tuned by varying an experimental parameter, for 
example the magnetic field in an Aharanov-Bohm interferometer with a quantum dot \cite{Kobayashi02}.

\begin{figure}[tb]
\begin{center}
\includegraphics[width=0.6\textwidth]{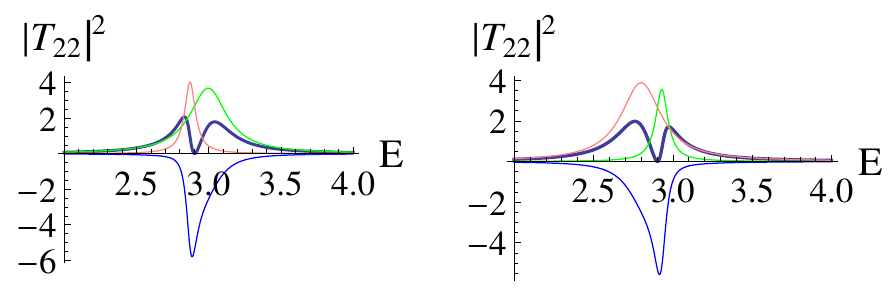}
\end{center}
%\begin{minipage}[b]{12pc}
\caption{Interference term (blue), 
the moduli squared of the two pole terms (green and pink) and $|T_{22}|^2$ (thick line) {\it vs.} energy
for $f=f_{\rm EP}+0.2$ (left) and $f=f_{\rm EP}-0.2$ (right).  }                  
%\end{minipage}
\end{figure}

Common to all illustrations is the ubiquitous asymmetry inherent in Fano-Feshbach resonances.
It is caused by the different widths of the pole terms and by the interference term,
most pronounced in Fig.4 even though there the interference
term is small in comparison with the contributions of the two pole terms.

\begin{figure}[htbp]
\begin{center}
\includegraphics[width=0.6\textwidth]{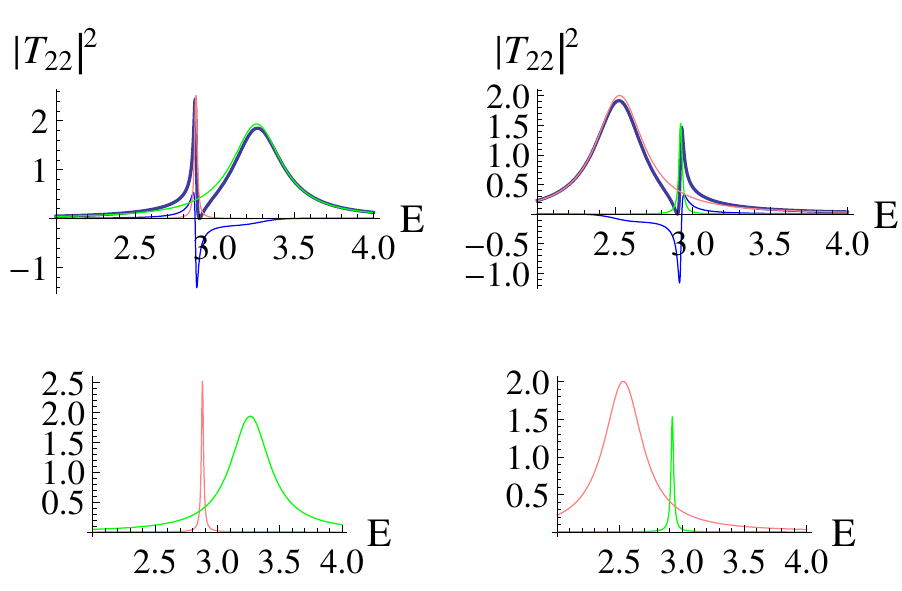}
\end{center}
%\begin{minipage}[b]{12pc}
\caption{Interference term (blue), 
moduli squared of the two pole terms (green and pink) and $|T_{22}|^2$ (thick line) {\it vs.} energy
for $f=f_{\rm EP}+1$ (left) and $f=f_{\rm EP}-1$ (right). For clarity the contributions of the individual pole terms
are displayed in the bottom row.  }                  
%\end{minipage}
\end{figure}

\begin{figure}[htbp]
\begin{center}
\includegraphics[width=0.5\textwidth]{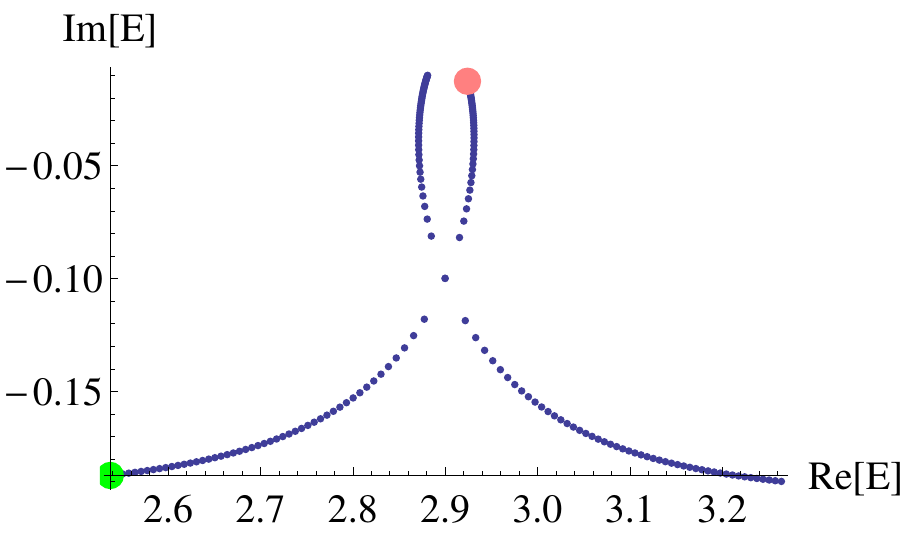}
\end{center}
%\begin{minipage}[b]{12pc}
\caption{Trajectories in the complex energy plane when $f$ is varied in the interval [-1,1].
The EP occurs at $E=2.9-\rmi \,0.099$ where the trajectories bounce off each other. The red and green dots
indicate the starting points for $f=-1$.  }                  
%\end{minipage}
\end{figure}

The energy trajectories illustrated in Fig.5 for $-1< f < +1$
clearly indicate the changing of the widths when $f$ is varied over the specified range
(we recall that the trajectories are in perfect agreement over the range considered irrespective of the full
(\ref{fep}) or the reduced problem (\ref{ham})).
When the interference becomes
larger while approaching the EP (Fig.2 and Fig.3), the asymmetry is always discernible and caused to an increasing
extent by the interference term
while the pole terms become more and more equal until they coalesce at the EP (Fig.1).
There the most remarkable feature are the two (unequal) peaks, where neither of them can
be associated with a single resonance pole. In all cases considered the interference term
produces the zero scattering at an energy between the two peaks or -- for different parameters -- a minimum.

\section{Summary and Conclusions}
Starting from the fact that in scattering systems EPs give rise to a second order pole in the Green's function,
in addition to the usual first order pole, we have investigated the impact on the shape of the scattering cross section as one approaches the EP. We have demonstrated that the merging  of the two poles results in manifestly asymmetric line profiles closely resembling Fano-Feshbach resonances. We have demonstrated that it is the interference term of the poles which generates the asymmetric line profiles.
In particular, at a certain energy it can produce a zero in the cross section, i.~e. with no scattering at all, or 
just a minimum.
We have used a simple classical system of two coupled oscillators with damping driven by an external force to 
highlight the effect of the interference term, but our findings apply quite generally to 
scattering cross sections close to an EP in open quantum systems.

 The present paper has established a close link between the appearance of Fano-Feshbach-like line profiles and 
the occurrence of EPs in parameter space. The review article \cite{Miroshnichenko10}
describes numerous experiments where in recent years asymmetric Fano-Feshbach-like line profiles have 
been observed, among them experiments with the interaction of light with structured matter, matter-wave scattering
in Bose-Einstein condensates, light scattering  by nanoparticles, and plasmonic micro cavities, to name just a few.  
From our findings it can be conjectured that all these asymmetric line profiles may be related to the occurrence of
EPs in the parameter spaces of the corresponding experiments. To verify this conjecture in actual experiments remains
a challenge for the analysis of such experiments.
\ack
We gratefully acknowledge the support from the National Institute
for Theoretical Physics (NITheP), Western Cape, South Africa. GW expresses his
gratitude to the Department of Physics of the University of Stellenbosch for
kind hospitality while this manuscript was prepared.
\\

\section*{References}
\bibliographystyle{unsrt}
%\bibliography{literature}

%\begin{thebibliography}{10}
%\end{thebibliography}

\end{document}